%% file: paper22.tex
\documentclass[fleqn,usenatbib]{mnras}

\usepackage[T1]{fontenc}

\DeclareRobustCommand{\VAN}[3]{#2}
\let\VANthebibliography\thebibliography
\def\thebibliography{\DeclareRobustCommand{\VAN}[3]{##3}\VANthebibliography}

\usepackage{graphicx}	
\usepackage{amsmath}	
\usepackage{amssymb}	
\usepackage{bm}
\usepackage{csquotes}
\usepackage[shortlabels]{enumitem}
\usepackage{newtxtext,newtxmath}

\def\draftversion{1} 
\input{header.tex}

\newcommand{\vrot}{$v_{\rm rot,g}$\xspace}
\newcommand{\vdisp}{$\sigma_{\rm g}$\xspace}
\newcommand{\vsigma}{$v_{\rm rot,g}/\sigma_{\rm g}$\xspace}

\title[Phases of star formation in the MW]{VINTERGATAN IV: Cosmic phases of star formation in Milky Way-like galaxies}

\author[Segovia Otero et al.]{
Álvaro Segovia Otero,\thanks{alvaro.segovia@astro.lu.se}
Florent Renaud,
and Oscar Agertz
\\
Department of Astronomy and Theoretical Physics, Lund Observatory, Box 43, SE-221 00 Lund, Sweden\\
}

\pubyear{2015}

\begin{document}
\label{firstpage}
\pagerange{\pageref{firstpage}--\pageref{lastpage}}
\maketitle

\begin{abstract}
The star formation history of a galaxy is modulated by a plethora of internal processes and environmental conditions. The details of how these evolve and couple together is not fully understood yet. In this work, we study the effects that galaxy mergers and morphological transformations have on setting different modes of star formation at galactic scales and across cosmic time. We monitor the global properties of \vintergatan, a 20 pc resolution cosmological zoom-in simulation of a Milky Way-type galaxy. Between redshifts 1 and 5, we find that major mergers trigger multiple starburst episodes, corresponding to a tenfold drop of the gas depletion time down to 100 Myr. Bursty star formation is enabled by the emergence of a galactic disc, when the rotational velocity of gas starts to dominate over its velocity dispersion. Coherent motions of gas then outweigh disordered ones, such that the galaxy responds to merger-induced forcings by redistributing large amounts of gas towards high densities. As a result, the overall star formation rate is enhanced with an associated decrease in the depletion time. Before redshift 5, mergers are expected to be even more frequent. However, a more turbulent interstellar medium, is incapable of reacting in such a collective manner so as to spark rapid star formation. Thus, a constant long depletion time of 1 Gyr is kept, along with a low, but gradually increasing star formation rate. After the last major merger at redshift 1, \vintergatan spends the next 8 Gyr evolving secularly. It has a settled and adiabatically growing disc, and a constant star formation rate with gas depletion times of 1-2 Gyr. Our results are compatible with the observed rapid transition between different modes of star formation when galaxies leave the main sequence.

\end{abstract}




\section{Introduction}

Understanding star formation is a multi-scale, multi-physics problem and one of the main challenges in modern astrophysics. Over the last couple of decades, catalogues of extragalactic sources based on CO line fluxes, millimeter dust photometry, and far-infrared dust emission have enable the study of star formation properties across cosmic time \citep[e.g.][and references therein]{tacconi20,saintonge22}. One of the major statistical conclusions stemming from such a wealth of data is that most star-forming galaxies up to {$z \approx 6$} lie on the so called ``main sequence'' of star formation \citep[hereafter MS,][]{noeske07,elbaz07,elbaz11,genzel10,whitaker14,tacconi13,saintonge17}. The MS correlates star formation rates (SFR) and stellar masses ({$M_{\star}$}), and evolves as a function of cosmic time. Galaxies on the MS formed stars at higher rates (per unit of stellar mass) in the earlier Universe, being affected by different gas accretion and assembly histories \citep{whitaker12}. Hence, it is remarkable that at a given stellar mass and redshift, the scatter of the MS is small \citep[{$0.2-0.4$} dex, see][and references therein]{speagle14}.

MS galaxies are mostly blue in rest-frame {$U-V$} bands \citep{brinchmann04,vilellarojo21}, commonly host rotationally-supported discs with {$v_{\rm rot,gas}/\sigma_{\rm gas} > 1$} \citep{wisnioski15,jones21,frasermcklevie21}, and feature S{\'e}rsic indices of {$n \approx 1-2$} \citep{wuyts11,lang14,osborne20}. Their gas reservoirs change with time following the normalisation of the MS, with gas fractions ranging from 10\% in the local Universe \citep{leroy08} up to 60\% at {$z \approx 4$} \citep{tacconi10}. This indicates a strong connection between gas and SFRs, well-known for most MS galaxies, but not fully understood in detail.

The Kennicutt-Schmidt relation \citep[hereafter KS,][]{schmidt59,kennicutt98} is an empirical result showing the dependency between surface densities of gas and SFR as {$\Sigma_{\rm SFR} \propto \Sigma_{\rm gas}^n$}. Discs in this diagram broadly follow the so-called canonical KS relation, with {$n \approx 1.4$} \citep[][]{kennicutt12}. The depletion time, {$\tau_{\rm dep} = \Sigma_{\rm gas} / \Sigma_{\rm SFR} = M_{\rm gas}/\rm SFR$}, provides a timescale for star formation: discs along the MS spanning masses of {$9 < M_{\star}/{\rm M_{\odot}} < 12$} have, on average, {$\tau_{\rm dep} \approx 0.5-2$} Gyr \citep[e.g.][]{wang22}, with a moderate decrease of up to {$z \approx 3$} \citep{genzel15,tacconi18}. This means that smooth continuous accretion of gas is required to sustain star formation throughout the lifetime of a galaxy. 

Starburst galaxies are in stark contrast with this dominant mode of star formation, and are characterised as out-of-equilibrium systems with intense star formation activity on timescales of {$10-500$} Myr \citep{pan18,birkin21}. Either because of an enhanced {$\Sigma_{\rm SFR}$}, reduced {$\Sigma_{\rm gas}$}, or a combination of both, starburst galaxies make a separate trend in KS space with significant drops in \tdep \citep{daddi10b,rodighiero11}. They present enhanced SFRs per unit mass, but can also be found within the scatter of the MS \citep[][]{puglisi21,gomezguijarro22}. In the local Universe, they generally correspond to luminous and ultra-luminous infrared galaxies (LIRGs and ULIRGs) powered by ongoing wet major mergers \citep[][]{kim21}. However, moving to higher redshifts ({$z > 1$}), correlating short-lived bursts of star formation with galaxy interactions becomes a difficult task. Starbursts only contribute {$\sim$}5\% to the cosmic SFR density. While their fraction appears to be constant with time \citep{sargent12,schreiber15,hani20}, the merger rate is a continuously increasing function of redshift \citep[][see also \citealp{romano21,husko22} for merger rates of massive galaxies]{lotz11,duncan19}. Thus, it is of utter importance to investigate how the galactic gas reservoir responds to merger events at different cosmic epochs.

Large-scale cosmological simulations attempt to explain the mechanisms driving starbursts by reproducing the MS, the KS relation, and their outliers. They face two important issues. First, low spatial ({$\sim$}100 pc) and gas mass ({$> 10^6 \Msun$}) resolutions lead to a recovered starburst contribution of barely 1\% of the total SFR density \citep{vogelsberger14,sparre15}. Second, the choice of sub-grid models to describe the impact of feedback, turbulence, and star formation on the structure of the interstellar medium (ISM) significantly influences the outcome of simulations \citep{agertz13,hopkins13,sparre17}. With improved spatial ({$10-100$} pc) and gas mass ({$10^5 \Msun$}) resolutions, recent simulations retrieve realistic starburst fractions \citep[e.g.][]{montero19,nelson21}. Now, idealised isolated merger simulations \citep{hopkins18,moreno21} disentangle the building blocks of a standardised merger-induced starburst picture: interacting galaxies can exert torques that globally channel gas towards the centre, compress it into higher surface densities, and spark nuclear starbursts \citep{keel85}. With spatial and mass resolution on the order of parsecs and {$\sim$}{$10^3 \Msun$} respectively, they are also capable of tracking how the local turbulent and density structures of the ISM affect star formation. \citet{renaud14} and \citet{sparre16} register merger-driven SFR enhancements and tenfold declines in \tdep. \citet{renaud14} established the link between gravitational interactions and the local turbulence properties and gas probability density functions (PDFs) of the ISM. Compressive tides and turbulence translate the global merger-driven forcing into boosted galactic star formation over large volumes. Both nuclear and off-centered starbursts are reproduced from an excess of high density gas. Nevertheless, a consistent view including the full cosmological context of a galaxy is still missing \citep[specially at high redshift where high gas fractions blur the picture of this coupling, see e.g. ][]{perret14,fensch17}.

In this work we use \vintergatan, a 20 pc resolution cosmological zoom-in simulation of a Milky Way-like galaxy, to study how the evolution of the cosmological environment and properties of the ISM lead to different phases of star formation. The layout of the paper is the following: Section~\ref{sec:methods} briefly summarises the simulation setup; Section~\ref{sec:res} shows how the starburst activity is associated with both mergers and disc formation; Section~\ref{sec:conc} first lists the main conclusions and then discusses the implications of this work.

\begin{figure*}
	\includegraphics[width=\textwidth]{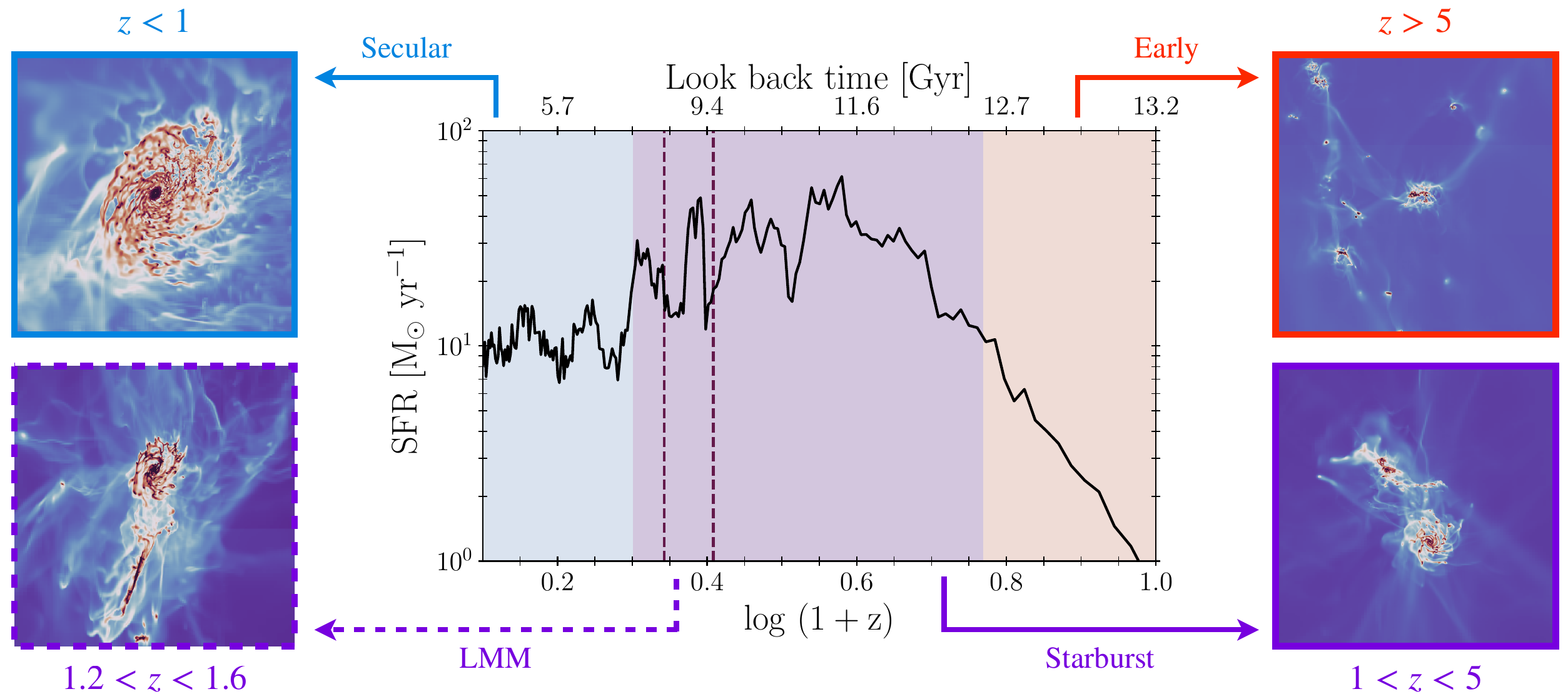}
    \caption{Star formation history of \vintergatan. The plot is colour-coded to highlight three epochs in chronological order (from right to left): \textit{Early} (orange), \textit{Starburst} (purple), and \textit{Secular} (blue). The vertical dashed lines mark the first pericenter passage ({$z \approx 1.6$}) and the final coalescence ({$z \approx 1.2$}) of the last major merger. An illustrative thumbnail 40 kpc wide shows the projected gas density at each evolutionary stage. During the \textit{Starburst} phase, as opposed to the \textit{Early} and \textit{Secular} ones, the SFR reaches several short-lived maxima above 40 {${\rm M}_{\odot} \ {\rm yr}^{-1}$}.}
    \label{fig:sfh}
\end{figure*}

\section{Method}
\label{sec:methods}
\defcitealias{vinter1}{Paper I}
\defcitealias{vinter2}{Paper II}
\defcitealias{vinter3}{Paper III}

This work uses the \vintergatan cosmological zoom-in simulation \citep[hereafter \citetalias{vinter1,vinter2,vinter3}]{vinter1,vinter2,vinter3}. A brief summary of the numerical recipe is provided below, with a more detailed description in \citetalias{vinter1}.

The simulation was run using \ramses \citep{teyssier02}, a hydrodynamic+{$N$}-body code with adaptive mesh refinement. Its initial conditions are the same as those of "m12i" from \citet{hopkins14}, i.e. a virial\footnote{This value is measured at {$\Delta = 200$} with respect to the mean cosmic background density.} radius of {$R_{200,m} = 334 \ {\rm kpc}$} and a virial mass of {$M_{200,m} = 1.3 \times 10^{12} \ {\rm M}_{\odot}$} at {$z = 0$}. Within a cosmological box of 85 Mpc containing {$512^3$} particles, a zoom-in technique \citep{hannabel11} is applied to the progenitor halo, achieving resolutions of {$3.5 \times 10^4 \ {\rm M}_{\odot}$} for dark matter, {$7070 \ {\rm M}_{\odot}$} for gas, and a physical resolution of {$\sim$}20 pc in the dense ISM. We stopped the simulation at {$z \approx 0.17$} (2.1 Gyr), due to computational costs and because the galaxy was not evolving significantly passed this redshift. The result is a Milky Way-type galaxy of {$\sim$}{$7 \times 10^{10} \ {\rm M_{\odot}}$} in stellar mass and {$\sim$}{$10^{10} \ {\rm M_{\odot}}$} in gas.

Star formation is modelled as a Poisson process, where star particles of {$10^4 \ {\rm M_{\odot}}$} are generated on a cell-by-cell basis following the law
\begin{equation}
    \Dot{\rho}_{\star} = \epsilon_{\rm ff} \dfrac{\rho_{\rm gas}}{t_{\rm ff}} \quad {\rm with} \quad \rho_{\rm gas} > 100 \ \mathrm{cm^{-3}} \quad {\rm and} \quad T_{\rm gas} < 100 \ \mathrm{K}.
    \label{eq:sf}
\end{equation}
Here, {$\Dot{\rho}_{\star}$} is the star formation rate density, {$t_{\rm ff}$} is the local free-fall time and {$\epsilon_{\rm ff}$} is the local star formation efficiency per free-fall time, parametrised according to \citet{padoan12}. These star particles represent individual stellar populations with stellar feedback processes including stellar winds, radiation pressure, type II, and type Ia supernovae \citep{agertz13,agertzkravtsov15,agertzkravtsov16}.

Cooling is metallicity-dependent and adopted from \citet{suthdopita93} for {$T > 10^4$} K, and \citet{rosenbergmas95} for {$T < 10^4$} K. An initial gas metallicity floor is set to 10{$^{-3}$} Z{$_{\odot}$} to reproduce the enrichment from unresolved population III stars \citep{wise2012}. Also accounted for is the additional gas heating from the ultraviolet background radiation field \citep{haardmadau96} assuming a reionization epoch at {$z = 8.5$}.

\section{Results}
\label{sec:res}

As demonstrated in \citetalias{vinter1}, \vintergatan reproduces key observational results on the formation of the Milky Way, including the chemical bimodality in [{$\alpha$}/Fe]-[Fe/H], the thin-thick disc dichotomy, and the similarity of the surface brightness and rotational velocity profiles with that of nearby disc galaxies. Therefore, our simulated galaxy sets realistic grounds for studies of the physics of star formation in Milky Way-like galaxies and their progenitors.

\subsection{Star formation history}
\label{sec:sfh}

Figure~\ref{fig:sfh} shows the star formation history of \vintergatan, calculated including all stars in the last snapshot ({$z \approx 0.2$}, 2.2 Gyr ago) within a sphere of radius 3{$R_{1/2}$}\footnote{{$R_{1/2}$} is computed for the most massive progenitor of \vintergatan for each snapshot, as the half-mass radius of stars younger than 100 Myr \citepalias[see][]{vinter2}.}, and arranged in age bins of 150 Myr. From right to left in Figure~\ref{fig:sfh}, we define three distinct phases:

\begin{enumerate}[a)]
    \item \textit{Early} ({$z>4.8$}, 12.7 Gyr ago): \vintergatan is slowly, but steadily increasing its SFR of a few {${\rm M}_{\odot} \ {\rm yr}^{-1}$}. Our small compact gas-rich galaxy is therefore smoothly building its stellar mass ({$R_{1/2} \approx 1$} kpc, {$M_{\star} < 10^{9} \Msun, f_{\rm gas} \gg 50\%$}). Mergers are frequent and prevent the development of an ordered morphology \citepalias{vinter2}. In spite of the elevated rate of galaxy interactions, no starburst activity exists at this stage.
    
    \item \textit{Starburst} ({$1<z<4.8$}, {$7.8-12.7$} Gyr ago): the galaxy transitions into a phase still merger-dominated, but much more active in terms of star formation. In the span of 5 Gyr, \vintergatan experiences four major merger events that give rise to SFR peaks of {$\sim$}{$40 \ {\rm M}_{\odot} \ {\rm yr}^{-1}$} \citepalias{vinter2}.
    
    \item \textit{Secular} ({$z<1$}, 7.8 Gyr ago): comparing the top left and bottom left thumbnails in Figure~\ref{fig:sfh} shows that after the last major merger (LMM, coallescence at {$z \approx 1.2$}), \vintergatan settles into a massive and extended disc ({$R_{1/2} \approx 4.8$} kpc, {$M_{\star} \approx 10^{11} \Msun$}, {$f_{\rm gas} \approx 10-20\%$}, \citetalias{vinter1}). Its SFR has moderate oscillations of {$\sim$}3 {${\rm M}_{\odot} \ {\rm yr}^{-1}$} when minor satellites accrete onto the main galaxy. Both {$M_{\star}$} and the SFR are compatible with that of observed MS discs at these redshifts \citep{whitaker12,speagle14}.  
\end{enumerate}

This first diagnostic suggests that mergers are a necessary but not sufficient condition to boost the SFR. Although the merger rate decreases with {$z$} \citep[as inferred statistically from observations, e.g.][and simulations, e.g. \citealt{fakhouri10}]{rodighiero11,schreiber15}, intense phases of star formation only occur between redshifts 1 and 5. Another key factor in the star formation history and overall mass assembly of galaxies is gas accretion. However, as shown in Figure 2 of \citetalias[][]{vinter2}, high inflow rates are mostly consequences of merger events and do not play a major role in enhancing the SFR.

Since the environment of \vintergatan in both the \textit{Early} and \textit{Starburst} epochs is similar in terms of interactions, we suspect that structural modifications in the ISM are responsible for the different modes of star formation. This is explained in Section~\ref{sec:vsigma}. 

\begin{figure}
	\includegraphics[width=\columnwidth]{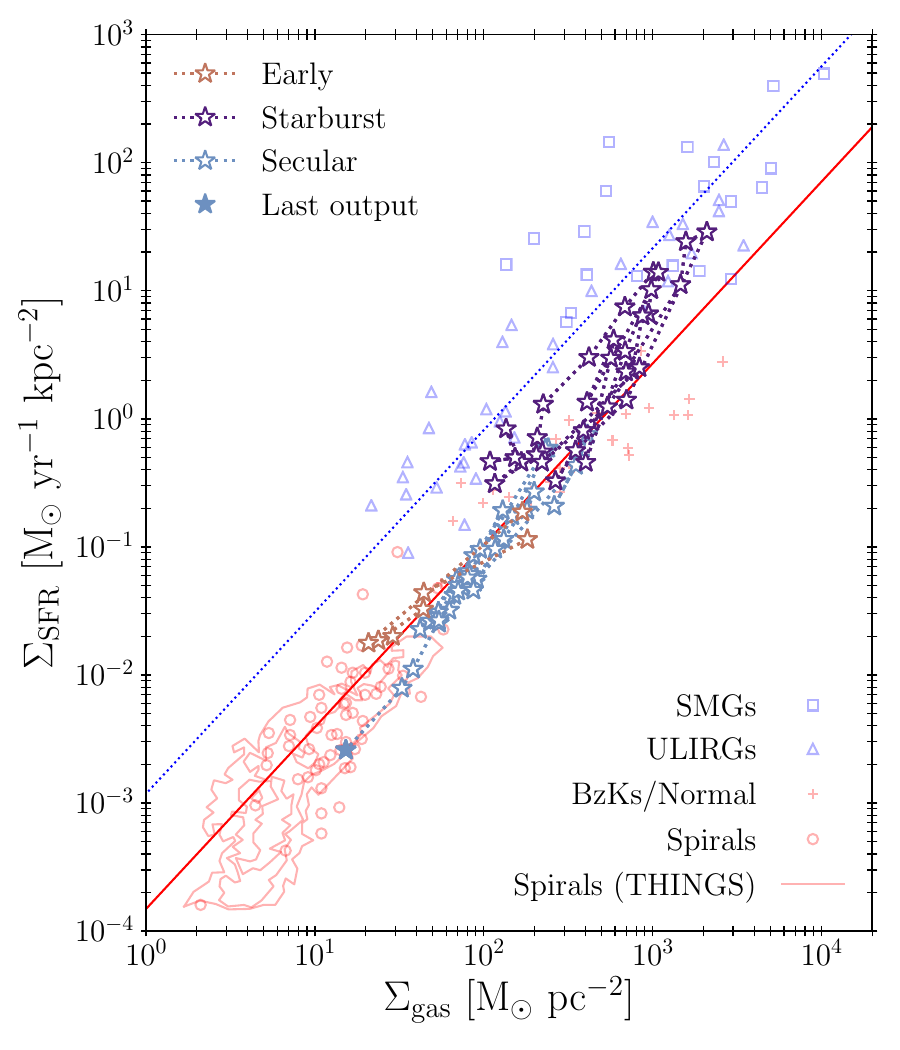}
    \caption{KS relation comparing observations of extragalactic sources (bottom right legend) with \vintergatan (top left legend). Observations comprise sub-millimeter galaxies at {$1.4 < z < 3.4$} \citep[SMGs,][]{bouche07,bothwell09}, ULIRGs at low redshift \citep{kennicutt98}, BzKs at {$z \approx 1.5$} \citep{daddi10a}, normal galaxies at {$1 < z < 2.3$} \citep{tacconi10}, spiral galaxies \citep{kennicutt98}, and spirals from the THINGS survey \citep{bigiel08}. Star markers indicate the location of \vintergatan on this plane at every output of the simulation, colour-coded according to the different phases identified in Figure~\ref{fig:sfh}. The first 7 outputs form the \textit{Early} phase, the following 33 form the \textit{Starburst} phase, and 36 the \textit{Secular} phase.} Along its lifetime, the galaxy transitions from the disc sequence (solid red line) and reaches the starburst sequence \citep[dotted blue line, Equation 2 in][]{daddi10b} during its \textit{Starburst} phase, before coming back to the disc sequence.
    \label{fig:ksd}
\end{figure}

\subsection{The Kennicutt-Schmidt relation}
\label{sec:ksd}

The evolution of a galaxy in the KS diagram evaluates to what degree it is undergoing a starburst episode, provided a control sample. Galaxies in Figure~\ref{fig:ksd} can be grouped in two categories: those represented by red markers along the canonical KS relation, also referred to as the ``disc sequence'' (solid red line), and observations displayed in blue around the ``starburst sequence'' (dotted blue line), shifted {$\sim$}1 dex above the disc sequence. High redshift gas-rich galaxies on the disc sequence have increased {$\Sigma_{\rm SFR}$} and {$\Sigma_{\rm gas}$} compared to local discs, with only a slight decrease in \tdep \citep{daddi10a}, whereas galaxies on the starburst sequence feature a significantly lower \tdep \citep{daddi10b,rodighiero11}.

For \vintergatan, {$\Sigma_{\rm gas}$} and {$\Sigma_{\rm SFR}$} are calculated on each simulation snapshot by accounting for the mass of cold gas ({$T < 10^4$} K) and stars younger than 100 Myr. Both quantities are computed within a sphere of radius 3{$R_{1/2}$}. The evolution of \vintergatan in Figure~\ref{fig:ksd} shows that the galaxy resides in the disc sequence in the \textit{Early} phase, jumps onto the starburst sequence during the \textit{Starburst} phase, and returns back to the disc sequence in the \textit{Secular} phase, consistent with nearby discs in the THINGS survey \citep{bigiel08}.

\subsection{Depletion time across cosmic time}
\label{sec:tdep}

\begin{figure}
	\includegraphics[width=\columnwidth]{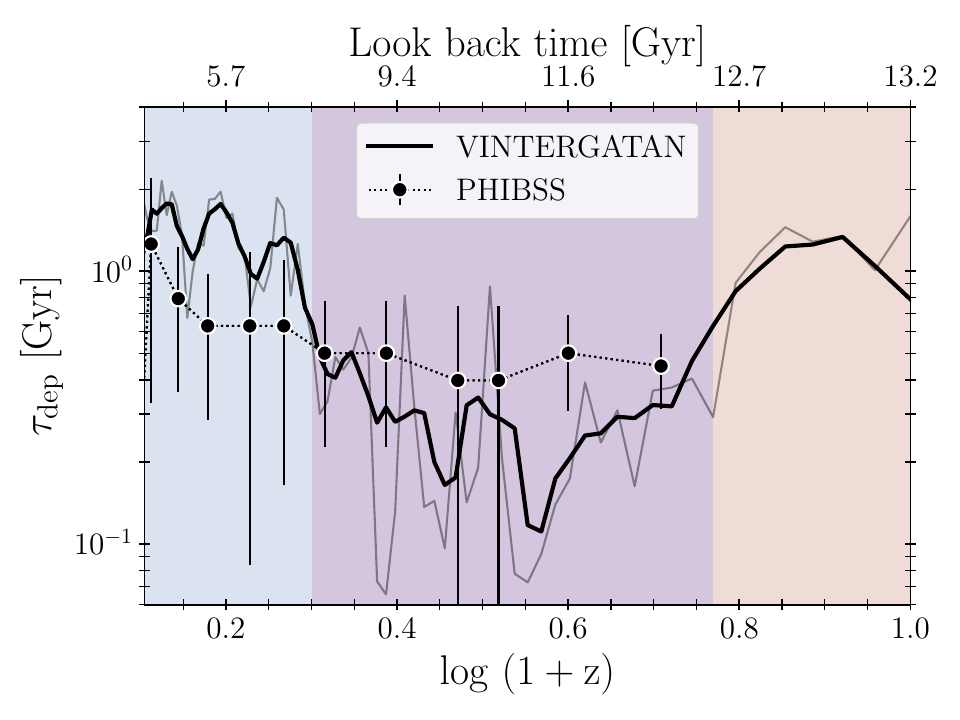}
    \caption{Cosmic evolution of \tdep in \vintergatan in solid grey, smoothed for clarity in black. We also show the median of the depletion times of galaxies in the PHIBSS survey \citep{tacconi18} divided in 30 redshift bins in logarithmic scale covering {$0.003 < z < 4.5$}. The error bars represent the median absolute deviation ({${\rm MAD} = 0.67449 \sigma$}, \citealt{muller00}). The coloured regions follow the same coding as in Figure~\ref{fig:sfh}. Overall, {$\tau_{\rm dep}$} drops by an order of magnitude in the \textit{Starburst} epoch with respect to the \textit{Secular} and \textit{Early} ones.}
    \label{fig:tdep}
\end{figure}

In Figure~\ref{fig:tdep}, we trace the evolution of \tdep in \vintergatan and compare it to observed data of star-forming galaxies from the PHIBSS survey \citep[][]{tacconi18}. In the \textit{Secular} phase (blue shaded region), the depletion time of our simulated galaxy varies between {$1-2$} Gyr. This is in agreement with observations of the nearby galaxies in \citet{leroy13}, but PHIBSS galaxies display higher SFRs at similar gas masses, leading to systematically lower \tdep.In the \textit{Starburst} epoch (comprising several major mergers, purple shaded region), \tdep repeatedly drops to 100 Myr. The general trend of our simulation is consistent with the PHIBSS data, but the latter shows a slight decline of a factor of {$\sim$}3 up to {$z \approx 4.5$}. The regularity of the PHIBSS curve is a consequence of stacking, which emphasises the statistically significant behaviour of a large sample of galaxies as a function of redshift. Stacking smooths out the fluctuations from individual galaxies like \vintergatan, with more abrupt dips in \tdep due to mergers.

One of the major sources of uncertainty in deriving \tdep from observations is the CO-to-H{$_{2}$} conversion factor. For instance, \aco translates the CO emission intensity into the total mass of H{$_{2}$}. \aco is observed to fluctuate with the galactic environment, e.g. between isolated disc galaxies and mergers \citep{bolatto13}. The underlying reasons are however debated. Based on predictions from simulations, \citet{renaud19a} propose a model to adjust \aco as a function of the stage of a merger, such that a reduced \aco is expected when the galaxy hosts a starburst. By applying this model to the \tdep of all the PHIBSS galaxies, we find that \tdep would change by a factor less than 2, at {$z>1$}. Nevertheless, this effect happens to be comparable to or even smaller than the scatter in \aco due to the use of different CO lines for each galaxy (see Appendix~\ref{sec:aco}).

A striking result from Figure~\ref{fig:tdep} is the long \tdep during the \textit{Early} phase, despite the merger activity. At this epoch, the galaxy forms stars with depletion times of {$\sim$}1 Gyr. This is ten times slower than in the \textit{Starburst} epoch. These changes in depletion time reflect modifications of the density structure of the ISM, as shown below.

\subsection{Gas density PDFs}
\label{sec:npdfs}

According to turbulence theory of isothermal supersonic gas, the density PDF of the ISM follows a log-normal functional form \citep{federrath10}. For isolated disc simulations, log-normal distributions provide a good fit, even with a non-isothermal ISM \citep{robertsonkravtsov08,renaud13}. The gas density PDF is sensitive to the volume over which it is measured. This is particularly important in cosmological simulations, where inflows, outflows and incoming galaxies can modify the shape of the PDF, specially at its low-density end. This blurs the information on local turbulence properties conveyed by the PDF.

\begin{figure}
	\includegraphics[width=\columnwidth]{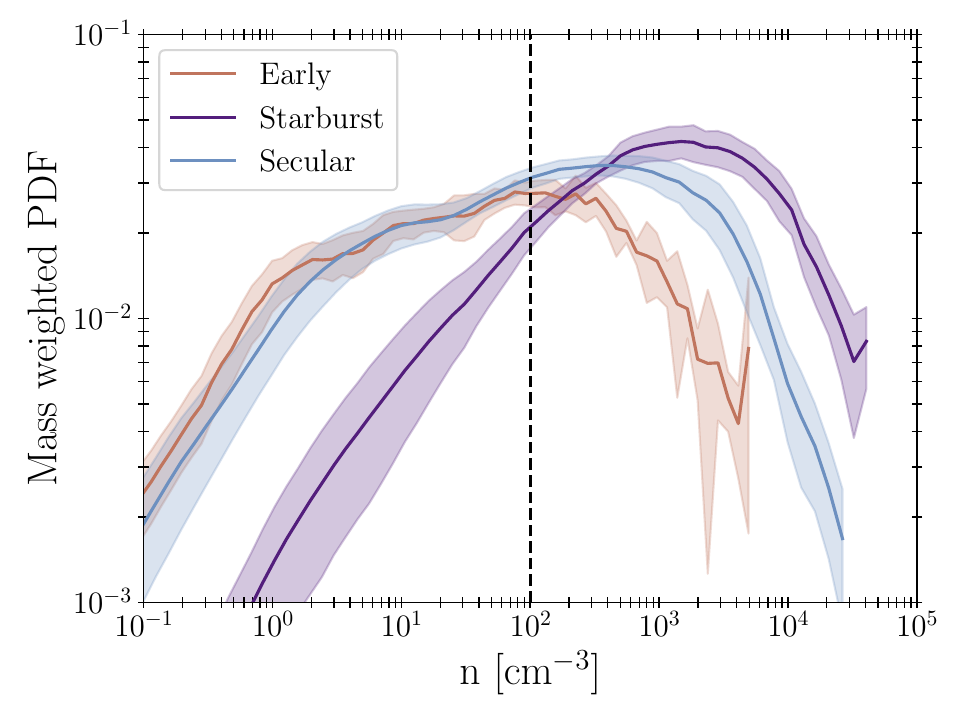}
    \caption{Gas density PDFs of \vintergatan. For each phase, the stacked PDFs show the normalized distribution of mass-weighted densities, averaged over the number of snapshots per cosmic epoch as indicated in Figure~\ref{fig:ksd}.} The shaded regions delimit the {$\sigma/3$} region, and the vertical dashed line marks the threshold for star formation (Equation~\ref{eq:sf}). The \textit{Starburst} phase (when the galaxy experiences rapid star formation episodes, i.e. low \tdep caused by major mergers) shows an excess of gas at high densities.
    \label{fig:npdfs}
\end{figure}

We focus on understanding whether the evolution of \tdep in our three phases corresponds to variations in the shape of the density PDFs (Figure~\ref{fig:npdfs}). In comparison with the \textit{Secular} phase, the PDF in the \textit{Starburst} epoch shows an excess of gas at high densities. During this epoch, large amounts of gas are compressed into dense star-forming states, leading to rapid star formation events, i.e. short \tdep. In idealised simulation of mergers, the excess of dense gas and the reduction of the depletion time have been associated with tidal compression \citep{renaud09,renaud14,renaud19b}. This is also the case in cosmological context (see Renaud et al. in prep.). This is why \vintergatan jumps to the starburst regime in the KS plane, rather than moving along the canonical relation. We stress that this conclusion is independent of any uncertainty on {$\alpha_{\rm CO}$} (Figure \ref{fig:ksd}).

Conversely, the gas in the \textit{Early} phase does not reach such dense states in spite of \vintergatan interacting with neighbour galaxies. This suggests a shift in the structure of the ISM between the \textit{Early} and \textit{Starburst} phases.

\subsection{Disc assembly}
\label{sec:vsigma}

\begin{figure}
	\includegraphics[width=\columnwidth]{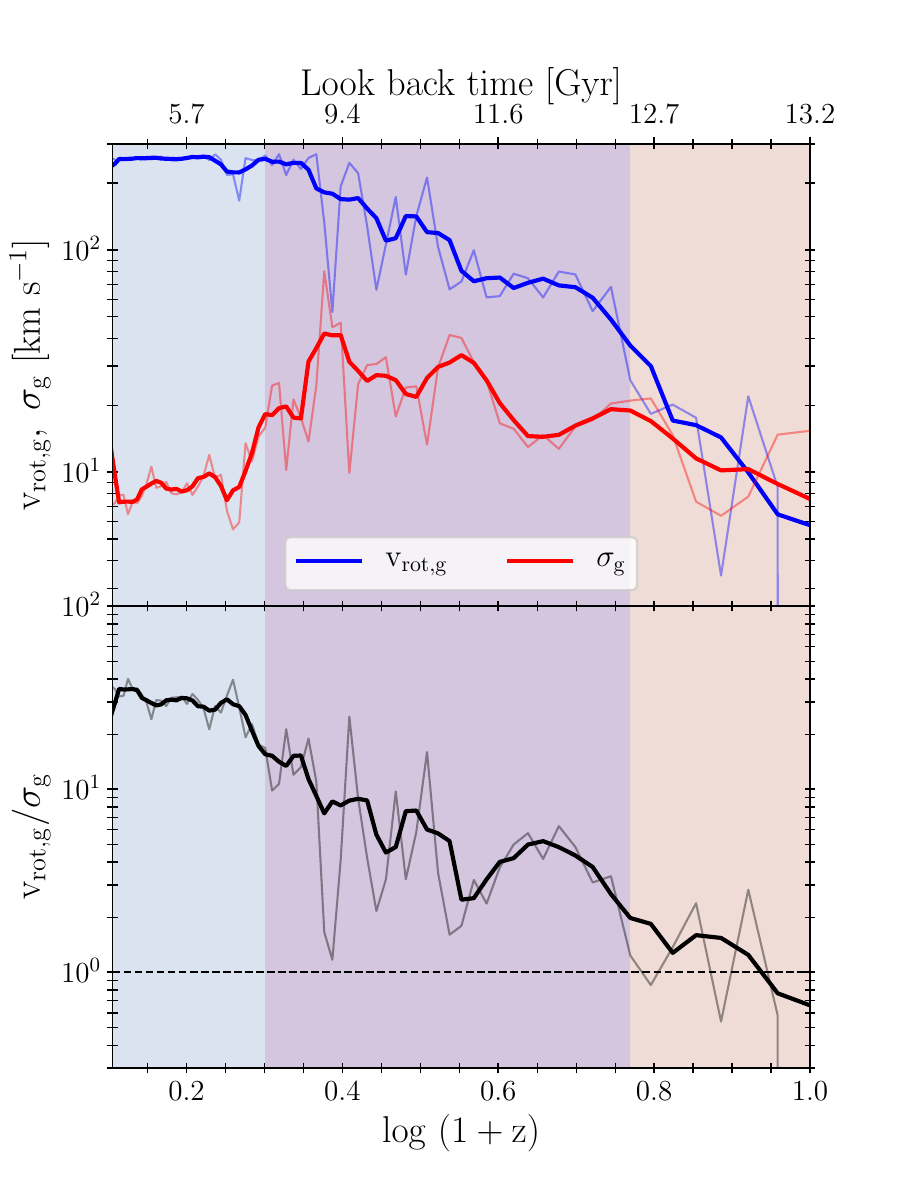}
    \caption{\textit{Top:} Rotational velocity \vrot (blue) and velocity dispersion \vdisp (red) of the cold gas as a function of redshift in \vintergatan; \textit{Bottom:} \vsigma ratio, used to parametrise disc assembly with cosmic time. All three curves have been smoothed in the same way as in Figure~\ref{fig:tdep}, and the shaded regions are coloured following the same code as in Figure~\ref{fig:sfh}. The transition at {$\log \left(1+z\right) \approx 0.75$} (i.e. {$z \approx 4.8$}, 12.5 Gyr ago) initiates the formation of a galactic disc, coeval with an excess of dense gas causing a drop in \tdep and the onset of starbursts (Figures~\ref{fig:tdep} and \ref{fig:npdfs}). A more rotationally-supported disc settles after the LMM with {$v_{\rm rot,g}/\sigma_{\rm g} > 10$} at {$\log \left(1+z\right) \approx 0.3$} (i.e. {$z \approx 1.2$}, 8.5 Gyr ago).}
    \label{fig:vsigma}
\end{figure}

Gas kinematics is a robust tracer of disc assembly, with high-quality observations characterising the dynamical state of a large sample of galaxies across cosmic time, \citep[e.g.][]{wisnioski19,rizzo21}. Studies in this field commonly use the gas rotational velocity \vrot to quantify organised circular motions, and the velocity dispersions \vdisp for the turbulent behaviour of the gas. One can consider that the disc is in place when {$v_{\rm rot,g}/\sigma_{\rm g} > 1$}, i.e. when circular motions dominate turbulent ones.

For every snapshot, we compute the rotation axis of the main progenitor galaxy as the total angular momentum vector of the cold gas ({$T < 10^4$} K) within {$3R_{1/2}$}. We then split the enclosed volume into a grid of (100 pc){$^{3}$} bins and, to ensure statistical significance, we evaluate the \vsigma ratio only in those bins that contain more than 10 cold gas cells. For \vrot, we use the modulus of the tangential velocity in cylindrical coordinates. We calculate \vdisp at 100 pc scale via {$\sigma_{\rm g}^2 = (\sigma_{\rm x}^2 + \sigma_{\rm y}^2 + \sigma_{\rm z}^2)/3$}, where {$\sigma_{i}$} represents the standard deviation of each component of the total gas velocity {$\vv{v}_{\rm g}$}.

Each point shown in Figure \ref{fig:vsigma} corresponds to the median value of these bins for each output of the simulation. Throughout the \textit{Secular} phase, we see a steep decline of {$\sigma_{\rm g}$} with decreasing redshift from {$\sim$}40 km s{$^{-1}$} to {$<$}10 km s{$^{-1}$}, along with a rather constant rotational velocity above {$\sim$}200 km s{$^{-1}$}. This is a strong indication of disc settling \citepalias{vinter3}, consistent with the kinematic downsizing picture of \citet{kassin12} and \citet{simons17}. In the \textit{Starburst} phase, the presence of ongoing interactions (and the associated starburst and stellar feedback) sustain a high \vdisp with a remarkable increase in \vrot with decreasing redshift. Despite such a violent environment, the \vsigma ratio stays above unity.

In the earliest epochs ({$z>4.8$}), the morphology of the galaxy is complex. Computing the angular momentum vector of this object leads to fluctuating orientations of the rotation axis and therefore substantially different values of the rotational velocities. This, together with velocity dispersions of comparable amplitude as \vrot, indicates that coherent and disordered motions are of the same order of magnitude ({$v_{\rm rot,g} \approx \sigma_{\rm g}$}). According to the kinematic downsizing scenario \citep{kassin12,simons17}, the formation of a galactic disc for a Milky Way progenitor ({$M_{\star} \sim 10^9$} M{$_{\odot}$}) is unlikely before {$z \approx 5$}. Mergers, gas accretion from counter-rotating streams, stellar feedback and violent disc instabilities during this period allude to high-redshift galaxies with large \vsigma as rare objects. Yet, this has been questioned by the recent detections of \citet{neeleman20,rizzo20,fraternali21,lelli21}. These massive ({$M_{\star} \gtrsim 10^{10}$} M{$_{\odot}$}), dusty, starburst discs at {$z > 4$} have rotational velocities of the order of hundreds km s{$^{-1}$}, and 10 times lower velocity dispersion. Long-lived discs at high redshift can also be reproduced in simulations if galaxies are massive and isolated enough, where filamentary accretion of gas with high angular momentum is co-planar and aligned with the disc \citep{dekel20a,tamfal21,kretschmer22}. Despite \vintergatan being an order of magnitude less massive than the aforementioned objects and hosting more modest rotational velocities at the same epoch, our results are in agreement with these studies.

In summary, regardless of how violent the environment of \vintergatan is, reduced depletion times indicating starbursts ({$\sim$}100 Myr) are only achieved once a galactic disc is in place. In our simulation, this starts at {$z \approx 4.8$}, when {$v_{\rm rot,g} / \sigma_{\rm g} > 1$}. Later ($z<1$), in the absence of mergers to compress the ISM and rapidly enhance the SFR, our galaxy grows in size by a factor of {$\sim$}5 \citepalias{vinter1}, and its depletion time increases to {$\sim$}few Gyr.

\section{Discussion \& Conclusion}
\label{sec:conc}

Using the \vintergatan cosmological zoom-in simulation \citepalias{vinter1,vinter2,vinter3}, we identify three phases of star formation along the evolution of a Milky Way-type galaxy, consistent with the star formation relations derived from observations:

\begin{enumerate}[a)]
    \item Early on ({$z > 4.8$}, 12.7 Gyr ago), the gas depletion time in \vintergatan is long ({$\sim$}1 Gyr) despite the ubiquity of mergers in this epoch. Interactions prevent the galaxy from developing a well-ordered morphology and kinematics, maintaining a turbulence level of {$\sigma_{\rm g} \sim 10-20$} km s{$^{-1}$} in the cold gas. Hence, our small ({$R_{1/2} \approx 400$} pc), gas-rich ({$f_{\rm g} \gg 50\%$}) galaxy has a non-bursty star formation, with rates below {$\sim$}{$10 \Msun$} yr{$^{-1}$}. This epoch corresponds to the building of the stellar population at high-[{$\alpha$}/Fe] and low, but quickly rising, metallicity ([Fe/H] {$<-0.8$}), which possess halo-like kinematics \citepalias{vinter2}.
    
    \item At redshifts {$1 < z < 4.8$}, the kinematics of the cold gas indicate that the structure of the ISM shifts towards coherent motions dominating over turbulent ones ({$v_{\rm rot,g} > \sigma_{\rm g}$}). Then, during interaction and merger events, gas is compressed on galactic scales, creating an excess at high densities and a tenfold drop of the depletion time ({$\sim$}100 Myr). \vintergatan jumps to the starburst sequence in the KS plane \citep{daddi10b}, and is also consistent with outliers observed above the MS \citep{rodighiero11,tacconi18} with peaks of SFR {$\gtrsim 40 \Msun$} yr{$^{-1}$}. With a small disc in place and a consequent rapid growth of the galaxy (due to both accreted material and starburst activity), the population at high-[{$\alpha$}/Fe] now reaches high metallicities ([Fe/H] {$\approx 0.7$}). This is known as the kinematically-hot ``thick disc'' sequence \citepalias{vinter2}.
    
    \item Starbursts come to an end after the last major merger ({$z < 1$}, 7.8 Gyr ago), with depletion times rising back to {$1-2$} Gyr. This event marks a crucial point in the simulation, when the population at low-[{$\alpha$}/Fe], or ``thin disc'' sequence, forms simultaneously from its low and high metallicity ends \citepalias[][]{vinter3}. In the absence of merger-driven perturbations, the ISM does not maintain its excess of dense gas and the galaxy transitions back to the disc regime in the KS plane \citep{daddi10a,kennicutt12}. The galaxy settles into a massive ({$M_{\star} \approx 10^{11} \Msun$}), extended ({$R_{1/2} \approx 3$} kpc), and rotationally-supported ({$v_{\rm rot,g} / \sigma_{\rm g} > 10$}) disc, with a relatively quiescent SFR {$\approx 10 \Msun$} yr{$^{-1}$}. 
    This characterises the secular phase of star formation for the last 8 Gyr, placing \vintergatan on the  main-sequence \citep{whitaker12,speagle14,schreiber15}.
\end{enumerate}

Different modes of star formation have also been identified in previous simulations, but without a description of their evolution across cosmic time \citep[i.e. without cosmological context,][]{renaud14,renaud19b,li22}. Nevertheless, when mimicking high-redshift conditions, idealised simulations of gas-rich, clumpy galaxy mergers report no significant enhancement of star formation \citep{perret14, fensch17}. A possible interpretation for such an inefficiency is that the intrinsically high velocity dispersion of the gas in gas-rich discs ($\approx 40 \kms$) saturates during interactions, and thus cannot lead to further compression \citep{fensch17}. However, similar levels of turbulence are found in \vintergatan at {$1 < z < 5$}, yet allowing merger-triggered starbursts. Therefore, the discrepancy seen in the star formation response to mergers likely originates from the cosmological context. We suspect that repeated interactions (only present in cosmological setups) could inhibit the formation of massive clumps, or tidally disrupt them, modifying the structure of the galaxies between idealised and cosmological runs.

A similar outcome is also found in the idealised merger simulations of \citet{hopkins13}, with {$f_{\rm gas} = 50 \%$}, but with massive clumps rapidly dispersed with the adopted implementation of stellar feedback. Using the same feedback model as in \vintergatan, simulations of isolated discs show that gas fractions $\gtrsim 20\%$ allow for the formation of long-lived massive clumps \citep{renaud21,floor22}. Yet, the absence of such clumps in \vintergatan indicates that the morphology of the ISM could be a decisive factor for merger-driven starbursts, even after the formation of the disc.

This confirms conclusions from recent infrared observations of galaxies \citep{hogan22} and the analysis of metal-poor stellar populations in the Milky Way \citep{conroy22}, which have highlighted the role of disc morphologies and kinematics in influencing the star formation properties along the assembly of galaxies. Future surveys of high redshift sources \citep[JWST,][]{evans22} will help disentangling the complex interplay of mechanisms at stake, and the their evolution across cosmic time.

\section*{Acknowledgements}

ASO thanks Eric Andersson for insightful discussions and the anonymous referee for their comments that helped improve this manuscript. ASO, FR, and OA acknowledge support from the Knut and Alice Wallenberg Foundation and the Royal Physiographic Society of Lund. ASO and OA are supported by grant 2019-04659 from the Swedish Research Council. This work used the COSMA Data Centric system at Durham University, operated by the Institute for Computational Cosmology on behalf of the STFC DiRAC HPC Facility (www.dirac.ac.uk). This equipment was funded by a BIS National E-infrastructure capital grant ST/K00042X/1, DiRAC Operations grant ST/K003267/1 and Durham University. DiRAC is part of the National E-Infrastructure.

\section*{Data Availability}
 
The data underlying this article will be shared on reasonable request to the corresponding author.



\bibliographystyle{mnras}





\appendix
\section{Change in \tdep with \aco}
\label{sec:aco}

Galaxy interactions can potentially lead to short \tdep and episodes of enhanced star formation. Feedback injected subsequently can lead to reduced \aco values \citep[e.g.][]{narayanan11, renaud19a}. To quantify how this affects \tdep in the PHIBSS galaxies, we apply the relation from Figure 5 in \citet{renaud19a}, derived from simulations of mergers in starbursting phase:
\begin{equation}
    \alpha'_{\rm CO} = 1.33 \log (t_{\rm dep}/{\rm Myr}) + 0.13,
    \label{eq:acoFR}
\end{equation}
and then recompute the gas mass as {$M'_{\rm gas} =  \alpha'_{\rm CO} M_{\rm gas}/\alpha_{\rm CO}$}. Here \aco is derived from the PHIBSS catalogue, assuming that the CO 1-0 transition was used for every galaxy. We finally obtain the modified depletion time {$\tau'_{\rm dep} = M'_{\rm gas}/{\rm SFR}$}.

Furthermore, while the CO 1-0 line is favoured in the local Universe, observations of high redshift galaxies rely on higher CO transitions, implying different {$\alpha_{\rm CO}$} conversion factors. To compare the importance of using different transitions to that of a {$\tau_{\rm dep}$}-dependent {$\alpha_{\rm CO}$}, we compute another {$\tau_{\rm dep}$}, derived from the {$\alpha_{\rm CO}$} of the CO 3-2 transition, according to Equation 2 of \citet{tacconi18}. Without information on the merger phase nor on the CO transition used for each individual galaxy in PHIBSS, we apply this method to all the PHIBSS galaxies. 

\begin{figure}
	\includegraphics[width=\columnwidth]{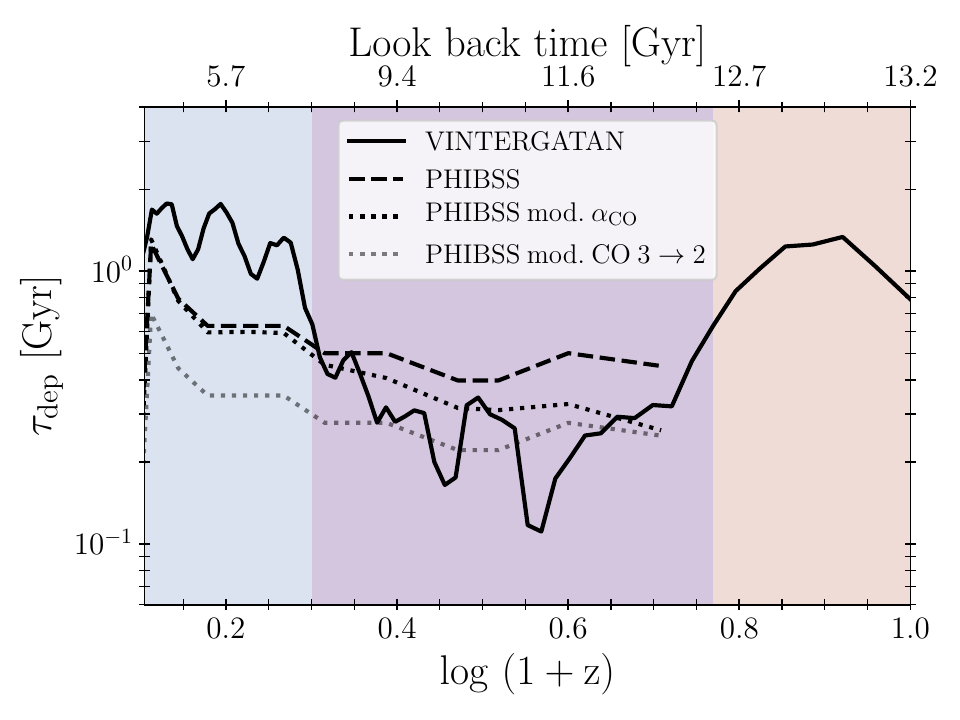}
    \caption{Gas depletion times from \vintergatan, compared to three estimates for the PHIBSS galaxies: (\emph{i}) the raw values from the catalogue (as in Figure \ref{fig:tdep}), (\emph{ii}) with an \aco modified to account for the starburst activity (Equation~\ref{eq:acoFR}, see \citealt{renaud19a}), and (\emph{iii}) with an \aco assuming the use of the CO 3-2 transition for all galaxies.}
    \label{fig:aco}
\end{figure}

Figure \ref{fig:aco} shows the evolution of the depletion times in \vintergatan and the PHIBSS galaxies (as in Figure \ref{fig:tdep}), and adds the comparison with the two corrections described above. Accounting for a {$\tau_{\rm dep}$}-dependent modification of {$\alpha_{\rm CO}$} (dotted black line) provides a better agreement between the observations and \vintergatan in the \textit{Starburst} stage. A similar effect is found when modifying \aco for the CO 3-2 transition (dotted grey line), although the use of this transition for all galaxies implies an over-correction at low redshift.

In conclusion, modifying \aco to account for the effects of merger-driven starbursts reconciles the depletion time of \vintergatan with that of the PHIBSS galaxies at high redshifts.

\bsp	
\label{lastpage}
\end{document}

%% file: header.tex
\usepackage{amssymb,latexsym,graphicx,natbib,eufrak,times,amsmath,xspace,ifthen}

\ifthenelse{ \draftversion > 0 }{
} {
  
}

\ifthenelse{\equal{\draftversion}{1}}{
  \usepackage{xcolor}
  \newcommand{\sep}[1]{\par\begin{color}[rgb]{0,0.4,0}\begin{center}\hrule\end{center}\end{color}\par} 
  \newcommand{\todo}[1]{\begin{color}{red}\ \ifthenelse{\equal{#1}{}} {$\bullet\bullet\bullet$} {$\bullet$\ #1 $\bullet$}\end{color}} 
  \newcommand{\idea}[1]{\begin{color}[rgb]{0,0.4,0}\textit{#1}\end{color}} 
  \newcommand{\sk}[1]{\begin{color}[rgb]{0.6,0,0.6}#1\end{color}} 
  \newcommand{\toc}{\par\begin{color}[rgb]{0.6,0,0.6}\begin{center}\hrule\vspace{0.5mm}\begingroup\small\let\cleardoublepage\relax\let\clearpage\relax\mytoc\endgroup\vspace{0.5mm}\hrule\end{center}\end{color}\par} 

  }{
  \newsavebox{\trashcan}

  \newcommand{\sep}[1]{}
  \newcommand{\todo}[1]{}
  \newcommand{\idea}[1]{}
  \newcommand{\sk}[1]{}
  \newcommand{\toc}{}

  }
 \makeatletter\newcommand\mytoc{\@starttoc{toc}}\makeatother 
\long\def\symbolfootnote[#1]#2{\begingroup%
\def\thefootnote{\fnsymbol{footnote}}\footnote[#1]{#2}\endgroup} 

\newcommand{\bb}[1]{\ifmmode \mbox{\boldmath $ #1$} \else  \mbox{\boldmath $#1$} \fi}

\newcommand{\U}[1]{\ensuremath{\mathrm{~#1}}}     

\newcommand{\Msun}{\U{M}_{\odot}}

\newcommand{\kms}{\U{km\ s^{-1}}}

\newcommand{\aco}{\ensuremath{\alpha_\mathrm{CO}}\xspace}     
\newcommand{\tdep}{\ensuremath{\tau_\mathrm{dep}}\xspace}        

\newcommand{\ramses}{{\small RAMSES}\xspace}

\newcommand{\vintergatan}{{\small VINTERGATAN}\xspace}






%% file: paper22.bbl
\begin{thebibliography}{}
\makeatletter
\relax
\def\mn@urlcharsother{\let\do\@makeother \do\$\do\&\do\#\do\^\do\_\do\%\do\~}
\def\mn@doi{\begingroup\mn@urlcharsother \@ifnextchar [ {\mn@doi@}
  {\mn@doi@[]}}
\def\mn@doi@[#1]#2{\def\@tempa{#1}\ifx\@tempa\@empty \href
  {http://dx.doi.org/#2} {doi:#2}\else \href {http://dx.doi.org/#2} {#1}\fi
  \endgroup}
\def\mn@eprint#1#2{\mn@eprint@#1:#2::\@nil}
\def\mn@eprint@arXiv#1{\href {http://arxiv.org/abs/#1} {{\tt arXiv:#1}}}
\def\mn@eprint@dblp#1{\href {http://dblp.uni-trier.de/rec/bibtex/#1.xml}
  {dblp:#1}}
\def\mn@eprint@#1:#2:#3:#4\@nil{\def\@tempa {#1}\def\@tempb {#2}\def\@tempc
  {#3}\ifx \@tempc \@empty \let \@tempc \@tempb \let \@tempb \@tempa \fi \ifx
  \@tempb \@empty \def\@tempb {arXiv}\fi \@ifundefined
  {mn@eprint@\@tempb}{\@tempb:\@tempc}{\expandafter \expandafter \csname
  mn@eprint@\@tempb\endcsname \expandafter{\@tempc}}}

\bibitem[\protect\citeauthoryear{{Agertz} \& {Kravtsov}}{{Agertz} \&
  {Kravtsov}}{2015}]{agertzkravtsov15}
{Agertz} O.,  {Kravtsov} A.~V.,  2015, \mn@doi [\apj]
  {10.1088/0004-637X/804/1/18}, \href
  {https://ui.adsabs.harvard.edu/abs/2015ApJ...804...18A} {804, 18}

\bibitem[\protect\citeauthoryear{{Agertz} \& {Kravtsov}}{{Agertz} \&
  {Kravtsov}}{2016}]{agertzkravtsov16}
{Agertz} O.,  {Kravtsov} A.~V.,  2016, \mn@doi [\apj]
  {10.3847/0004-637X/824/2/79}, \href
  {https://ui.adsabs.harvard.edu/abs/2016ApJ...824...79A} {824, 79}

\bibitem[\protect\citeauthoryear{{Agertz}, {Kravtsov}, {Leitner}  \&
  {Gnedin}}{{Agertz} et~al.}{2013}]{agertz13}
{Agertz} O.,  {Kravtsov} A.~V.,  {Leitner} S.~N.,   {Gnedin} N.~Y.,  2013,
  \mn@doi [\apj] {10.1088/0004-637X/770/1/25}, \href
  {https://ui.adsabs.harvard.edu/abs/2013ApJ...770...25A} {770, 25}

\bibitem[\protect\citeauthoryear{{Agertz} et~al.,}{{Agertz}
  et~al.}{2021}]{vinter1}
{Agertz} O.,  et~al., 2021, \mn@doi [\mnras] {10.1093/mnras/stab322}, \href
  {https://ui.adsabs.harvard.edu/abs/2021MNRAS.503.5826A} {503, 5826}

\bibitem[\protect\citeauthoryear{{Bigiel}, {Leroy}, {Walter}, {Brinks}, {de
  Blok}, {Madore}  \& {Thornley}}{{Bigiel} et~al.}{2008}]{bigiel08}
{Bigiel} F.,  {Leroy} A.,  {Walter} F.,  {Brinks} E.,  {de Blok} W.~J.~G.,
  {Madore} B.,   {Thornley} M.~D.,  2008, \mn@doi [\aj]
  {10.1088/0004-6256/136/6/2846}, \href
  {https://ui.adsabs.harvard.edu/abs/2008AJ....136.2846B} {136, 2846}

\bibitem[\protect\citeauthoryear{{Birkin} et~al.,}{{Birkin}
  et~al.}{2021}]{birkin21}
{Birkin} J.~E.,  et~al., 2021, \mn@doi [\mnras] {10.1093/mnras/staa3862}, \href
  {https://ui.adsabs.harvard.edu/abs/2021MNRAS.501.3926B} {501, 3926}

\bibitem[\protect\citeauthoryear{{Bolatto}, {Wolfire}  \& {Leroy}}{{Bolatto}
  et~al.}{2013}]{bolatto13}
{Bolatto} A.~D.,  {Wolfire} M.,   {Leroy} A.~K.,  2013, \mn@doi [\araa]
  {10.1146/annurev-astro-082812-140944}, \href
  {https://ui.adsabs.harvard.edu/abs/2013ARA&A..51..207B} {51, 207}

\bibitem[\protect\citeauthoryear{{Bothwell}, {Kennicutt}  \& {Lee}}{{Bothwell}
  et~al.}{2009}]{bothwell09}
{Bothwell} M.~S.,  {Kennicutt} R.~C.,   {Lee} J.~C.,  2009, \mn@doi [\mnras]
  {10.1111/j.1365-2966.2009.15471.x}, \href
  {https://ui.adsabs.harvard.edu/abs/2009MNRAS.400..154B} {400, 154}

\bibitem[\protect\citeauthoryear{{Bouch{\'e}} et~al.,}{{Bouch{\'e}}
  et~al.}{2007}]{bouche07}
{Bouch{\'e}} N.,  et~al., 2007, \mn@doi [\apj] {10.1086/522221}, \href
  {https://ui.adsabs.harvard.edu/abs/2007ApJ...671..303B} {671, 303}

\bibitem[\protect\citeauthoryear{{Brinchmann}, {Charlot}, {White}, {Tremonti},
  {Kauffmann}, {Heckman}  \& {Brinkmann}}{{Brinchmann}
  et~al.}{2004}]{brinchmann04}
{Brinchmann} J.,  {Charlot} S.,  {White} S.~D.~M.,  {Tremonti} C.,  {Kauffmann}
  G.,  {Heckman} T.,   {Brinkmann} J.,  2004, \mn@doi [\mnras]
  {10.1111/j.1365-2966.2004.07881.x}, \href
  {https://ui.adsabs.harvard.edu/abs/2004MNRAS.351.1151B} {351, 1151}

\bibitem[\protect\citeauthoryear{{Conroy} et~al.,}{{Conroy}
  et~al.}{2022}]{conroy22}
{Conroy} C.,  et~al., 2022, arXiv e-prints, \href
  {https://ui.adsabs.harvard.edu/abs/2022arXiv220402989C} {p. arXiv:2204.02989}

\bibitem[\protect\citeauthoryear{{Daddi} et~al.,}{{Daddi}
  et~al.}{2010a}]{daddi10a}
{Daddi} E.,  et~al., 2010a, \mn@doi [\apj] {10.1088/0004-637X/713/1/686}, \href
  {https://ui.adsabs.harvard.edu/abs/2010ApJ...713..686D} {713, 686}

\bibitem[\protect\citeauthoryear{{Daddi} et~al.,}{{Daddi}
  et~al.}{2010b}]{daddi10b}
{Daddi} E.,  et~al., 2010b, \mn@doi [\apjl] {10.1088/2041-8205/714/1/L118},
  \href {https://ui.adsabs.harvard.edu/abs/2010ApJ...714L.118D} {714, L118}

\bibitem[\protect\citeauthoryear{{Dekel}, {Ginzburg}, {Jiang}, {Freundlich},
  {Lapiner}, {Ceverino}  \& {Primack}}{{Dekel} et~al.}{2020}]{dekel20a}
{Dekel} A.,  {Ginzburg} O.,  {Jiang} F.,  {Freundlich} J.,  {Lapiner} S.,
  {Ceverino} D.,   {Primack} J.,  2020, \mn@doi [\mnras]
  {10.1093/mnras/staa470}, \href
  {https://ui.adsabs.harvard.edu/abs/2020MNRAS.493.4126D} {493, 4126}

\bibitem[\protect\citeauthoryear{{Duncan} et~al.,}{{Duncan}
  et~al.}{2019}]{duncan19}
{Duncan} K.,  et~al., 2019, \mn@doi [\apj] {10.3847/1538-4357/ab148a}, \href
  {https://ui.adsabs.harvard.edu/abs/2019ApJ...876..110D} {876, 110}

\bibitem[\protect\citeauthoryear{{Elbaz} et~al.,}{{Elbaz}
  et~al.}{2007}]{elbaz07}
{Elbaz} D.,  et~al., 2007, \mn@doi [\aap] {10.1051/0004-6361:20077525}, \href
  {https://ui.adsabs.harvard.edu/abs/2007A&A...468...33E} {468, 33}

\bibitem[\protect\citeauthoryear{{Elbaz} et~al.,}{{Elbaz}
  et~al.}{2011}]{elbaz11}
{Elbaz} D.,  et~al., 2011, \mn@doi [\aap] {10.1051/0004-6361/201117239}, \href
  {https://ui.adsabs.harvard.edu/abs/2011A&A...533A.119E} {533, A119}

\bibitem[\protect\citeauthoryear{{Evans}, {Fattahi}, {Deason}  \&
  {Frenk}}{{Evans} et~al.}{2022}]{evans22}
{Evans} T.~A.,  {Fattahi} A.,  {Deason} A.~J.,   {Frenk} C.~S.,  2022, arXiv
  e-prints, \href {https://ui.adsabs.harvard.edu/abs/2022arXiv220401794E} {p.
  arXiv:2204.01794}

\bibitem[\protect\citeauthoryear{{Fakhouri}, {Ma}  \&
  {Boylan-Kolchin}}{{Fakhouri} et~al.}{2010}]{fakhouri10}
{Fakhouri} O.,  {Ma} C.-P.,   {Boylan-Kolchin} M.,  2010, \mn@doi [\mnras]
  {10.1111/j.1365-2966.2010.16859.x}, \href
  {https://ui.adsabs.harvard.edu/abs/2010MNRAS.406.2267F} {406, 2267}

\bibitem[\protect\citeauthoryear{{Federrath}, {Roman-Duval}, {Klessen},
  {Schmidt}  \& {Mac Low}}{{Federrath} et~al.}{2010}]{federrath10}
{Federrath} C.,  {Roman-Duval} J.,  {Klessen} R.~S.,  {Schmidt} W.,   {Mac Low}
  M.~M.,  2010, \mn@doi [\aap] {10.1051/0004-6361/200912437}, \href
  {https://ui.adsabs.harvard.edu/abs/2010A&A...512A..81F} {512, A81}

\bibitem[\protect\citeauthoryear{{Fensch} et~al.,}{{Fensch}
  et~al.}{2017}]{fensch17}
{Fensch} J.,  et~al., 2017, \mn@doi [\mnras] {10.1093/mnras/stw2920}, \href
  {https://ui.adsabs.harvard.edu/abs/2017MNRAS.465.1934F} {465, 1934}

\bibitem[\protect\citeauthoryear{{Fraser-McKelvie} et~al.,}{{Fraser-McKelvie}
  et~al.}{2021}]{frasermcklevie21}
{Fraser-McKelvie} A.,  et~al., 2021, \mn@doi [\mnras] {10.1093/mnras/stab573},
  \href {https://ui.adsabs.harvard.edu/abs/2021MNRAS.503.4992F} {503, 4992}

\bibitem[\protect\citeauthoryear{{Fraternali}, {Karim}, {Magnelli},
  {G{\'o}mez-Guijarro}, {Jim{\'e}nez-Andrade}  \& {Posses}}{{Fraternali}
  et~al.}{2021}]{fraternali21}
{Fraternali} F.,  {Karim} A.,  {Magnelli} B.,  {G{\'o}mez-Guijarro} C.,
  {Jim{\'e}nez-Andrade} E.~F.,   {Posses} A.~C.,  2021, \mn@doi [\aap]
  {10.1051/0004-6361/202039807}, \href
  {https://ui.adsabs.harvard.edu/abs/2021A&A...647A.194F} {647, A194}

\bibitem[\protect\citeauthoryear{{Genzel} et~al.,}{{Genzel}
  et~al.}{2010}]{genzel10}
{Genzel} R.,  et~al., 2010, \mn@doi [\mnras]
  {10.1111/j.1365-2966.2010.16969.x}, \href
  {https://ui.adsabs.harvard.edu/abs/2010MNRAS.407.2091G} {407, 2091}

\bibitem[\protect\citeauthoryear{{Genzel} et~al.,}{{Genzel}
  et~al.}{2015}]{genzel15}
{Genzel} R.,  et~al., 2015, \mn@doi [\apj] {10.1088/0004-637X/800/1/20}, \href
  {https://ui.adsabs.harvard.edu/abs/2015ApJ...800...20G} {800, 20}

\bibitem[\protect\citeauthoryear{{G{\'o}mez-Guijarro}
  et~al.,}{{G{\'o}mez-Guijarro} et~al.}{2022}]{gomezguijarro22}
{G{\'o}mez-Guijarro} C.,  et~al., 2022, arXiv e-prints, \href
  {https://ui.adsabs.harvard.edu/abs/2022arXiv220102633G} {p. arXiv:2201.02633}

\bibitem[\protect\citeauthoryear{{Haardt} \& {Madau}}{{Haardt} \&
  {Madau}}{1996}]{haardmadau96}
{Haardt} F.,  {Madau} P.,  1996, \mn@doi [\apj] {10.1086/177035}, \href
  {https://ui.adsabs.harvard.edu/abs/1996ApJ...461...20H} {461, 20}

\bibitem[\protect\citeauthoryear{{Hahn} \& {Abel}}{{Hahn} \&
  {Abel}}{2011}]{hannabel11}
{Hahn} O.,  {Abel} T.,  2011, \mn@doi [\mnras]
  {10.1111/j.1365-2966.2011.18820.x}, \href
  {https://ui.adsabs.harvard.edu/abs/2011MNRAS.415.2101H} {415, 2101}

\bibitem[\protect\citeauthoryear{{Hani}, {Gosain}, {Ellison}, {Patton}  \&
  {Torrey}}{{Hani} et~al.}{2020}]{hani20}
{Hani} M.~H.,  {Gosain} H.,  {Ellison} S.~L.,  {Patton} D.~R.,   {Torrey} P.,
  2020, \mn@doi [\mnras] {10.1093/mnras/staa459}, \href
  {https://ui.adsabs.harvard.edu/abs/2020MNRAS.493.3716H} {493, 3716}

\bibitem[\protect\citeauthoryear{{Hogan} et~al.,}{{Hogan}
  et~al.}{2022}]{hogan22}
{Hogan} L.,  et~al., 2022, arXiv e-prints, \href
  {https://ui.adsabs.harvard.edu/abs/2022arXiv220210576H} {p. arXiv:2202.10576}

\bibitem[\protect\citeauthoryear{{Hopkins}, {Cox}, {Hernquist}, {Narayanan},
  {Hayward}  \& {Murray}}{{Hopkins} et~al.}{2013}]{hopkins13}
{Hopkins} P.~F.,  {Cox} T.~J.,  {Hernquist} L.,  {Narayanan} D.,  {Hayward}
  C.~C.,   {Murray} N.,  2013, \mn@doi [\mnras] {10.1093/mnras/stt017}, \href
  {https://ui.adsabs.harvard.edu/abs/2013MNRAS.430.1901H} {430, 1901}

\bibitem[\protect\citeauthoryear{{Hopkins}, {Kere{\v{s}}}, {O{\~n}orbe},
  {Faucher-Gigu{\`e}re}, {Quataert}, {Murray}  \& {Bullock}}{{Hopkins}
  et~al.}{2014}]{hopkins14}
{Hopkins} P.~F.,  {Kere{\v{s}}} D.,  {O{\~n}orbe} J.,  {Faucher-Gigu{\`e}re}
  C.-A.,  {Quataert} E.,  {Murray} N.,   {Bullock} J.~S.,  2014, \mn@doi
  [\mnras] {10.1093/mnras/stu1738}, \href
  {https://ui.adsabs.harvard.edu/abs/2014MNRAS.445..581H} {445, 581}

\bibitem[\protect\citeauthoryear{{Hopkins} et~al.,}{{Hopkins}
  et~al.}{2018}]{hopkins18}
{Hopkins} P.~F.,  et~al., 2018, \mn@doi [\mnras] {10.1093/mnras/sty1690}, \href
  {https://ui.adsabs.harvard.edu/abs/2018MNRAS.480..800H} {480, 800}

\bibitem[\protect\citeauthoryear{{Hu{\v{s}}ko}, {Lacey}  \&
  {Baugh}}{{Hu{\v{s}}ko} et~al.}{2022}]{husko22}
{Hu{\v{s}}ko} F.,  {Lacey} C.~G.,   {Baugh} C.~M.,  2022, \mn@doi [\mnras]
  {10.1093/mnras/stab3324}, \href
  {https://ui.adsabs.harvard.edu/abs/2022MNRAS.509.5918H} {509, 5918}

\bibitem[\protect\citeauthoryear{{Jones} et~al.,}{{Jones}
  et~al.}{2021}]{jones21}
{Jones} G.~C.,  et~al., 2021, \mn@doi [\mnras] {10.1093/mnras/stab2226}, \href
  {https://ui.adsabs.harvard.edu/abs/2021MNRAS.507.3540J} {507, 3540}

\bibitem[\protect\citeauthoryear{{Kassin} et~al.,}{{Kassin}
  et~al.}{2012}]{kassin12}
{Kassin} S.~A.,  et~al., 2012, \mn@doi [\apj] {10.1088/0004-637X/758/2/106},
  \href {https://ui.adsabs.harvard.edu/abs/2012ApJ...758..106K} {758, 106}

\bibitem[\protect\citeauthoryear{{Keel}, {Kennicutt}, {Hummel}  \& {van der
  Hulst}}{{Keel} et~al.}{1985}]{keel85}
{Keel} W.~C.,  {Kennicutt} R.~C. J.,  {Hummel} E.,   {van der Hulst} J.~M.,
  1985, \mn@doi [\aj] {10.1086/113779}, \href
  {https://ui.adsabs.harvard.edu/abs/1985AJ.....90..708K} {90, 708}

\bibitem[\protect\citeauthoryear{{Kennicutt}}{{Kennicutt}}{1998}]{kennicutt98}
{Kennicutt} Robert~C. J.,  1998, \mn@doi [\apj] {10.1086/305588}, \href
  {https://ui.adsabs.harvard.edu/abs/1998ApJ...498..541K} {498, 541}

\bibitem[\protect\citeauthoryear{{Kennicutt} \& {Evans}}{{Kennicutt} \&
  {Evans}}{2012}]{kennicutt12}
{Kennicutt} R.~C.,  {Evans} N.~J.,  2012, \mn@doi [\araa]
  {10.1146/annurev-astro-081811-125610}, \href
  {https://ui.adsabs.harvard.edu/abs/2012ARA&A..50..531K} {50, 531}

\bibitem[\protect\citeauthoryear{{Kim} et~al.,}{{Kim} et~al.}{2021}]{kim21}
{Kim} E.,  et~al., 2021, \mn@doi [\mnras] {10.1093/mnras/stab2090}, \href
  {https://ui.adsabs.harvard.edu/abs/2021MNRAS.507.3113K} {507, 3113}

\bibitem[\protect\citeauthoryear{{Kretschmer}, {Dekel}  \&
  {Teyssier}}{{Kretschmer} et~al.}{2022}]{kretschmer22}
{Kretschmer} M.,  {Dekel} A.,   {Teyssier} R.,  2022, \mn@doi [\mnras]
  {10.1093/mnras/stab3648}, \href
  {https://ui.adsabs.harvard.edu/abs/2022MNRAS.510.3266K} {510, 3266}

\bibitem[\protect\citeauthoryear{{Lang} et~al.,}{{Lang} et~al.}{2014}]{lang14}
{Lang} P.,  et~al., 2014, \mn@doi [\apj] {10.1088/0004-637X/788/1/11}, \href
  {https://ui.adsabs.harvard.edu/abs/2014ApJ...788...11L} {788, 11}

\bibitem[\protect\citeauthoryear{{Lelli}, {Di Teodoro}, {Fraternali}, {Man},
  {Zhang}, {De Breuck}, {Davis}  \& {Maiolino}}{{Lelli} et~al.}{2021}]{lelli21}
{Lelli} F.,  {Di Teodoro} E.~M.,  {Fraternali} F.,  {Man} A. W.~S.,  {Zhang}
  Z.-Y.,  {De Breuck} C.,  {Davis} T.~A.,   {Maiolino} R.,  2021, \mn@doi
  [Science] {10.1126/science.abc1893}, \href
  {https://ui.adsabs.harvard.edu/abs/2021Sci...371..713L} {371, 713}

\bibitem[\protect\citeauthoryear{{Leroy}, {Walter}, {Brinks}, {Bigiel}, {de
  Blok}, {Madore}  \& {Thornley}}{{Leroy} et~al.}{2008}]{leroy08}
{Leroy} A.~K.,  {Walter} F.,  {Brinks} E.,  {Bigiel} F.,  {de Blok} W.~J.~G.,
  {Madore} B.,   {Thornley} M.~D.,  2008, \mn@doi [\aj]
  {10.1088/0004-6256/136/6/2782}, \href
  {https://ui.adsabs.harvard.edu/abs/2008AJ....136.2782L} {136, 2782}

\bibitem[\protect\citeauthoryear{{Leroy} et~al.,}{{Leroy}
  et~al.}{2013}]{leroy13}
{Leroy} A.~K.,  et~al., 2013, \mn@doi [\aj] {10.1088/0004-6256/146/2/19}, \href
  {https://ui.adsabs.harvard.edu/abs/2013AJ....146...19L} {146, 19}

\bibitem[\protect\citeauthoryear{{Li}, {Vogelsberger}, {Bryan}, {Marinacci},
  {Sales}  \& {Torrey}}{{Li} et~al.}{2022}]{li22}
{Li} H.,  {Vogelsberger} M.,  {Bryan} G.~L.,  {Marinacci} F.,  {Sales} L.~V.,
  {Torrey} P.,  2022, \mn@doi [\mnras] {10.1093/mnras/stac1136}, \href
  {https://ui.adsabs.harvard.edu/abs/2022MNRAS.514..265L} {514, 265}

\bibitem[\protect\citeauthoryear{{Lotz}, {Jonsson}, {Cox}, {Croton}, {Primack},
  {Somerville}  \& {Stewart}}{{Lotz} et~al.}{2011}]{lotz11}
{Lotz} J.~M.,  {Jonsson} P.,  {Cox} T.~J.,  {Croton} D.,  {Primack} J.~R.,
  {Somerville} R.~S.,   {Stewart} K.,  2011, \mn@doi [\apj]
  {10.1088/0004-637X/742/2/103}, \href
  {https://ui.adsabs.harvard.edu/abs/2011ApJ...742..103L} {742, 103}

\bibitem[\protect\citeauthoryear{{Moreno} et~al.,}{{Moreno}
  et~al.}{2021}]{moreno21}
{Moreno} J.,  et~al., 2021, \mn@doi [\mnras] {10.1093/mnras/staa2952}, \href
  {https://ui.adsabs.harvard.edu/abs/2021MNRAS.503.3113M} {503, 3113}

\bibitem[\protect\citeauthoryear{M{\"u}ller}{M{\"u}ller}{2000}]{muller00}
M{\"u}ller J.,  2000, \mn@doi [Journal of Research of the National Institute of
  Standards and Technology] {10.6028/jres.105.044}, 105, 551

\bibitem[\protect\citeauthoryear{{Narayanan}, {Krumholz}, {Ostriker}  \&
  {Hernquist}}{{Narayanan} et~al.}{2011}]{narayanan11}
{Narayanan} D.,  {Krumholz} M.,  {Ostriker} E.~C.,   {Hernquist} L.,  2011,
  \mn@doi [\mnras] {10.1111/j.1365-2966.2011.19516.x}, \href
  {https://ui.adsabs.harvard.edu/abs/2011MNRAS.418..664N} {418, 664}

\bibitem[\protect\citeauthoryear{{Neeleman}, {Prochaska}, {Kanekar}  \&
  {Rafelski}}{{Neeleman} et~al.}{2020}]{neeleman20}
{Neeleman} M.,  {Prochaska} J.~X.,  {Kanekar} N.,   {Rafelski} M.,  2020,
  \mn@doi [\nat] {10.1038/s41586-020-2276-y}, \href
  {https://ui.adsabs.harvard.edu/abs/2020Natur.581..269N} {581, 269}

\bibitem[\protect\citeauthoryear{{Nelson} et~al.,}{{Nelson}
  et~al.}{2021}]{nelson21}
{Nelson} E.~J.,  et~al., 2021, \mn@doi [\mnras] {10.1093/mnras/stab2131}, \href
  {https://ui.adsabs.harvard.edu/abs/2021MNRAS.508..219N} {508, 219}

\bibitem[\protect\citeauthoryear{{Noeske} et~al.,}{{Noeske}
  et~al.}{2007}]{noeske07}
{Noeske} K.~G.,  et~al., 2007, \mn@doi [\apjl] {10.1086/517926}, \href
  {https://ui.adsabs.harvard.edu/abs/2007ApJ...660L..43N} {660, L43}

\bibitem[\protect\citeauthoryear{{Osborne} et~al.,}{{Osborne}
  et~al.}{2020}]{osborne20}
{Osborne} C.,  et~al., 2020, \mn@doi [\apj] {10.3847/1538-4357/abb5af}, \href
  {https://ui.adsabs.harvard.edu/abs/2020ApJ...902...77O} {902, 77}

\bibitem[\protect\citeauthoryear{{Padoan}, {Haugb{\o}lle}  \&
  {Nordlund}}{{Padoan} et~al.}{2012}]{padoan12}
{Padoan} P.,  {Haugb{\o}lle} T.,   {Nordlund} {\r{A}}.,  2012, \mn@doi [\apjl]
  {10.1088/2041-8205/759/2/L27}, \href
  {https://ui.adsabs.harvard.edu/abs/2012ApJ...759L..27P} {759, L27}

\bibitem[\protect\citeauthoryear{{Pan} et~al.,}{{Pan} et~al.}{2018}]{pan18}
{Pan} H.-A.,  et~al., 2018, \mn@doi [\apj] {10.3847/1538-4357/aaeb92}, \href
  {https://ui.adsabs.harvard.edu/abs/2018ApJ...868..132P} {868, 132}

\bibitem[\protect\citeauthoryear{{Perret}, {Renaud}, {Epinat}, {Amram},
  {Bournaud}, {Contini}, {Teyssier}  \& {Lambert}}{{Perret}
  et~al.}{2014}]{perret14}
{Perret} V.,  {Renaud} F.,  {Epinat} B.,  {Amram} P.,  {Bournaud} F.,
  {Contini} T.,  {Teyssier} R.,   {Lambert} J.~C.,  2014, \mn@doi [\aap]
  {10.1051/0004-6361/201322395}, \href
  {https://ui.adsabs.harvard.edu/abs/2014A&A...562A...1P} {562, A1}

\bibitem[\protect\citeauthoryear{{Puglisi} et~al.,}{{Puglisi}
  et~al.}{2021}]{puglisi21}
{Puglisi} A.,  et~al., 2021, \mn@doi [\mnras] {10.1093/mnras/stab2914}, \href
  {https://ui.adsabs.harvard.edu/abs/2021MNRAS.508.5217P} {508, 5217}

\bibitem[\protect\citeauthoryear{{Renaud}, {Boily}, {Naab}  \&
  {Theis}}{{Renaud} et~al.}{2009}]{renaud09}
{Renaud} F.,  {Boily} C.~M.,  {Naab} T.,   {Theis} C.,  2009, \mn@doi [\apj]
  {10.1088/0004-637X/706/1/67}, \href
  {https://ui.adsabs.harvard.edu/abs/2009ApJ...706...67R} {706, 67}

\bibitem[\protect\citeauthoryear{{Renaud} et~al.,}{{Renaud}
  et~al.}{2013}]{renaud13}
{Renaud} F.,  et~al., 2013, \mn@doi [\mnras] {10.1093/mnras/stt1698}, \href
  {https://ui.adsabs.harvard.edu/abs/2013MNRAS.436.1836R} {436, 1836}

\bibitem[\protect\citeauthoryear{{Renaud}, {Bournaud}, {Kraljic}  \&
  {Duc}}{{Renaud} et~al.}{2014}]{renaud14}
{Renaud} F.,  {Bournaud} F.,  {Kraljic} K.,   {Duc} P.~A.,  2014, \mn@doi
  [\mnras] {10.1093/mnrasl/slu050}, \href
  {https://ui.adsabs.harvard.edu/abs/2014MNRAS.442L..33R} {442, L33}

\bibitem[\protect\citeauthoryear{{Renaud}, {Bournaud}, {Daddi}  \&
  {Wei{\ss}}}{{Renaud} et~al.}{2019a}]{renaud19a}
{Renaud} F.,  {Bournaud} F.,  {Daddi} E.,   {Wei{\ss}} A.,  2019a, \mn@doi
  [\aap] {10.1051/0004-6361/201834397}, \href
  {https://ui.adsabs.harvard.edu/abs/2019A&A...621A.104R} {621, A104}

\bibitem[\protect\citeauthoryear{{Renaud}, {Bournaud}, {Agertz}, {Kraljic},
  {Schinnerer}, {Bolatto}, {Daddi}  \& {Hughes}}{{Renaud}
  et~al.}{2019b}]{renaud19b}
{Renaud} F.,  {Bournaud} F.,  {Agertz} O.,  {Kraljic} K.,  {Schinnerer} E.,
  {Bolatto} A.,  {Daddi} E.,   {Hughes} A.,  2019b, \mn@doi [\aap]
  {10.1051/0004-6361/201935222}, \href
  {https://ui.adsabs.harvard.edu/abs/2019A&A...625A..65R} {625, A65}

\bibitem[\protect\citeauthoryear{{Renaud}, {Agertz}, {Read}, {Ryde},
  {Andersson}, {Bensby}, {Rey}  \& {Feuillet}}{{Renaud}
  et~al.}{2021a}]{vinter2}
{Renaud} F.,  {Agertz} O.,  {Read} J.~I.,  {Ryde} N.,  {Andersson} E.~P.,
  {Bensby} T.,  {Rey} M.~P.,   {Feuillet} D.~K.,  2021a, \mn@doi [\mnras]
  {10.1093/mnras/stab250}, \href
  {https://ui.adsabs.harvard.edu/abs/2021MNRAS.503.5846R} {503, 5846}

\bibitem[\protect\citeauthoryear{{Renaud}, {Agertz}, {Andersson}, {Read},
  {Ryde}, {Bensby}, {Rey}  \& {Feuillet}}{{Renaud} et~al.}{2021b}]{vinter3}
{Renaud} F.,  {Agertz} O.,  {Andersson} E.~P.,  {Read} J.~I.,  {Ryde} N.,
  {Bensby} T.,  {Rey} M.~P.,   {Feuillet} D.~K.,  2021b, \mn@doi [\mnras]
  {10.1093/mnras/stab543}, \href
  {https://ui.adsabs.harvard.edu/abs/2021MNRAS.503.5868R} {503, 5868}

\bibitem[\protect\citeauthoryear{{Renaud}, {Romeo}  \& {Agertz}}{{Renaud}
  et~al.}{2021c}]{renaud21}
{Renaud} F.,  {Romeo} A.~B.,   {Agertz} O.,  2021c, \mn@doi [\mnras]
  {10.1093/mnras/stab2604}, \href
  {https://ui.adsabs.harvard.edu/abs/2021MNRAS.508..352R} {508, 352}

\bibitem[\protect\citeauthoryear{{Rizzo}, {Vegetti}, {Powell}, {Fraternali},
  {McKean}, {Stacey}  \& {White}}{{Rizzo} et~al.}{2020}]{rizzo20}
{Rizzo} F.,  {Vegetti} S.,  {Powell} D.,  {Fraternali} F.,  {McKean} J.~P.,
  {Stacey} H.~R.,   {White} S.~D.~M.,  2020, \mn@doi [\nat]
  {10.1038/s41586-020-2572-6}, \href
  {https://ui.adsabs.harvard.edu/abs/2020Natur.584..201R} {584, 201}

\bibitem[\protect\citeauthoryear{{Rizzo}, {Vegetti}, {Fraternali}, {Stacey}  \&
  {Powell}}{{Rizzo} et~al.}{2021}]{rizzo21}
{Rizzo} F.,  {Vegetti} S.,  {Fraternali} F.,  {Stacey} H.~R.,   {Powell} D.,
  2021, \mn@doi [\mnras] {10.1093/mnras/stab2295}, \href
  {https://ui.adsabs.harvard.edu/abs/2021MNRAS.507.3952R} {507, 3952}

\bibitem[\protect\citeauthoryear{{Robertson} \& {Kravtsov}}{{Robertson} \&
  {Kravtsov}}{2008}]{robertsonkravtsov08}
{Robertson} B.~E.,  {Kravtsov} A.~V.,  2008, \mn@doi [\apj] {10.1086/587796},
  \href {https://ui.adsabs.harvard.edu/abs/2008ApJ...680.1083R} {680, 1083}

\bibitem[\protect\citeauthoryear{{Rodighiero} et~al.,}{{Rodighiero}
  et~al.}{2011}]{rodighiero11}
{Rodighiero} G.,  et~al., 2011, \mn@doi [\apjl] {10.1088/2041-8205/739/2/L40},
  \href {https://ui.adsabs.harvard.edu/abs/2011ApJ...739L..40R} {739, L40}

\bibitem[\protect\citeauthoryear{{Rodr{\'\i}guez Montero}, {Dav{\'e}}, {Wild},
  {Angl{\'e}s-Alc{\'a}zar}  \& {Narayanan}}{{Rodr{\'\i}guez Montero}
  et~al.}{2019}]{montero19}
{Rodr{\'\i}guez Montero} F.,  {Dav{\'e}} R.,  {Wild} V.,
  {Angl{\'e}s-Alc{\'a}zar} D.,   {Narayanan} D.,  2019, \mn@doi [\mnras]
  {10.1093/mnras/stz2580}, \href
  {https://ui.adsabs.harvard.edu/abs/2019MNRAS.490.2139R} {490, 2139}

\bibitem[\protect\citeauthoryear{{Romano} et~al.,}{{Romano}
  et~al.}{2021}]{romano21}
{Romano} M.,  et~al., 2021, \mn@doi [\aap] {10.1051/0004-6361/202141306}, \href
  {https://ui.adsabs.harvard.edu/abs/2021A&A...653A.111R} {653, A111}

\bibitem[\protect\citeauthoryear{{Rosen} \& {Bregman}}{{Rosen} \&
  {Bregman}}{1995}]{rosenbergmas95}
{Rosen} A.,  {Bregman} J.~N.,  1995, \mn@doi [\apj] {10.1086/175303}, \href
  {https://ui.adsabs.harvard.edu/abs/1995ApJ...440..634R} {440, 634}

\bibitem[\protect\citeauthoryear{{Saintonge} \& {Catinella}}{{Saintonge} \&
  {Catinella}}{2022}]{saintonge22}
{Saintonge} A.,  {Catinella} B.,  2022, arXiv e-prints, \href
  {https://ui.adsabs.harvard.edu/abs/2022arXiv220200690S} {p. arXiv:2202.00690}

\bibitem[\protect\citeauthoryear{{Saintonge} et~al.,}{{Saintonge}
  et~al.}{2017}]{saintonge17}
{Saintonge} A.,  et~al., 2017, \mn@doi [\apjs] {10.3847/1538-4365/aa97e0},
  \href {https://ui.adsabs.harvard.edu/abs/2017ApJS..233...22S} {233, 22}

\bibitem[\protect\citeauthoryear{{Sargent}, {B{\'e}thermin}, {Daddi}  \&
  {Elbaz}}{{Sargent} et~al.}{2012}]{sargent12}
{Sargent} M.~T.,  {B{\'e}thermin} M.,  {Daddi} E.,   {Elbaz} D.,  2012, \mn@doi
  [\apjl] {10.1088/2041-8205/747/2/L31}, \href
  {https://ui.adsabs.harvard.edu/abs/2012ApJ...747L..31S} {747, L31}

\bibitem[\protect\citeauthoryear{{Schmidt}}{{Schmidt}}{1959}]{schmidt59}
{Schmidt} M.,  1959, \mn@doi [\apj] {10.1086/146614}, \href
  {https://ui.adsabs.harvard.edu/abs/1959ApJ...129..243S} {129, 243}

\bibitem[\protect\citeauthoryear{{Schreiber} et~al.,}{{Schreiber}
  et~al.}{2015}]{schreiber15}
{Schreiber} C.,  et~al., 2015, \mn@doi [\aap] {10.1051/0004-6361/201425017},
  \href {https://ui.adsabs.harvard.edu/abs/2015A&A...575A..74S} {575, A74}

\bibitem[\protect\citeauthoryear{{Simons} et~al.,}{{Simons}
  et~al.}{2017}]{simons17}
{Simons} R.~C.,  et~al., 2017, \mn@doi [\apj] {10.3847/1538-4357/aa740c}, \href
  {https://ui.adsabs.harvard.edu/abs/2017ApJ...843...46S} {843, 46}

\bibitem[\protect\citeauthoryear{{Sparre} \& {Springel}}{{Sparre} \&
  {Springel}}{2016}]{sparre16}
{Sparre} M.,  {Springel} V.,  2016, \mn@doi [\mnras] {10.1093/mnras/stw1793},
  \href {https://ui.adsabs.harvard.edu/abs/2016MNRAS.462.2418S} {462, 2418}

\bibitem[\protect\citeauthoryear{{Sparre} et~al.,}{{Sparre}
  et~al.}{2015}]{sparre15}
{Sparre} M.,  et~al., 2015, \mn@doi [\mnras] {10.1093/mnras/stu2713}, \href
  {https://ui.adsabs.harvard.edu/abs/2015MNRAS.447.3548S} {447, 3548}

\bibitem[\protect\citeauthoryear{{Sparre}, {Hayward}, {Feldmann},
  {Faucher-Gigu{\`e}re}, {Muratov}, {Kere{\v{s}}}  \& {Hopkins}}{{Sparre}
  et~al.}{2017}]{sparre17}
{Sparre} M.,  {Hayward} C.~C.,  {Feldmann} R.,  {Faucher-Gigu{\`e}re} C.-A.,
  {Muratov} A.~L.,  {Kere{\v{s}}} D.,   {Hopkins} P.~F.,  2017, \mn@doi
  [\mnras] {10.1093/mnras/stw3011}, \href
  {https://ui.adsabs.harvard.edu/abs/2017MNRAS.466...88S} {466, 88}

\bibitem[\protect\citeauthoryear{{Speagle}, {Steinhardt}, {Capak}  \&
  {Silverman}}{{Speagle} et~al.}{2014}]{speagle14}
{Speagle} J.~S.,  {Steinhardt} C.~L.,  {Capak} P.~L.,   {Silverman} J.~D.,
  2014, \mn@doi [\apjs] {10.1088/0067-0049/214/2/15}, \href
  {https://ui.adsabs.harvard.edu/abs/2014ApJS..214...15S} {214, 15}

\bibitem[\protect\citeauthoryear{{Sutherland} \& {Dopita}}{{Sutherland} \&
  {Dopita}}{1993}]{suthdopita93}
{Sutherland} R.~S.,  {Dopita} M.~A.,  1993, \mn@doi [\apjs] {10.1086/191823},
  \href {https://ui.adsabs.harvard.edu/abs/1993ApJS...88..253S} {88, 253}

\bibitem[\protect\citeauthoryear{{Tacconi} et~al.,}{{Tacconi}
  et~al.}{2010}]{tacconi10}
{Tacconi} L.~J.,  et~al., 2010, \mn@doi [\nat] {10.1038/nature08773}, \href
  {https://ui.adsabs.harvard.edu/abs/2010Natur.463..781T} {463, 781}

\bibitem[\protect\citeauthoryear{{Tacconi} et~al.,}{{Tacconi}
  et~al.}{2013}]{tacconi13}
{Tacconi} L.~J.,  et~al., 2013, \mn@doi [\apj] {10.1088/0004-637X/768/1/74},
  \href {https://ui.adsabs.harvard.edu/abs/2013ApJ...768...74T} {768, 74}

\bibitem[\protect\citeauthoryear{{Tacconi} et~al.,}{{Tacconi}
  et~al.}{2018}]{tacconi18}
{Tacconi} L.~J.,  et~al., 2018, \mn@doi [\apj] {10.3847/1538-4357/aaa4b4},
  \href {https://ui.adsabs.harvard.edu/abs/2018ApJ...853..179T} {853, 179}

\bibitem[\protect\citeauthoryear{{Tacconi}, {Genzel}  \& {Sternberg}}{{Tacconi}
  et~al.}{2020}]{tacconi20}
{Tacconi} L.~J.,  {Genzel} R.,   {Sternberg} A.,  2020, \mn@doi [\araa]
  {10.1146/annurev-astro-082812-141034}, \href
  {https://ui.adsabs.harvard.edu/abs/2020ARA&A..58..157T} {58, 157}

\bibitem[\protect\citeauthoryear{{Tamfal}, {Mayer}, {Quinn}, {Babul}, {Madau},
  {Capelo}, {Shen}  \& {Staub}}{{Tamfal} et~al.}{2021}]{tamfal21}
{Tamfal} T.,  {Mayer} L.,  {Quinn} T.~R.,  {Babul} A.,  {Madau} P.,  {Capelo}
  P.~R.,  {Shen} S.,   {Staub} M.,  2021, arXiv e-prints, \href
  {https://ui.adsabs.harvard.edu/abs/2021arXiv210611981T} {p. arXiv:2106.11981}

\bibitem[\protect\citeauthoryear{{Teyssier}}{{Teyssier}}{2002}]{teyssier02}
{Teyssier} R.,  2002, \mn@doi [\aap] {10.1051/0004-6361:20011817}, \href
  {https://ui.adsabs.harvard.edu/abs/2002A&A...385..337T} {385, 337}

\bibitem[\protect\citeauthoryear{{Vilella-Rojo} et~al.,}{{Vilella-Rojo}
  et~al.}{2021}]{vilellarojo21}
{Vilella-Rojo} G.,  et~al., 2021, \mn@doi [\aap] {10.1051/0004-6361/202039156},
  \href {https://ui.adsabs.harvard.edu/abs/2021A&A...650A..68V} {650, A68}

\bibitem[\protect\citeauthoryear{{Vogelsberger} et~al.,}{{Vogelsberger}
  et~al.}{2014}]{vogelsberger14}
{Vogelsberger} M.,  et~al., 2014, \mn@doi [\mnras] {10.1093/mnras/stu1536},
  \href {https://ui.adsabs.harvard.edu/abs/2014MNRAS.444.1518V} {444, 1518}

\bibitem[\protect\citeauthoryear{{Wang} et~al.,}{{Wang} et~al.}{2022}]{wang22}
{Wang} T.-M.,  et~al., 2022, arXiv e-prints, \href
  {https://ui.adsabs.harvard.edu/abs/2022arXiv220112070W} {p. arXiv:2201.12070}

\bibitem[\protect\citeauthoryear{{Whitaker}, {van Dokkum}, {Brammer}  \&
  {Franx}}{{Whitaker} et~al.}{2012}]{whitaker12}
{Whitaker} K.~E.,  {van Dokkum} P.~G.,  {Brammer} G.,   {Franx} M.,  2012,
  \mn@doi [\apjl] {10.1088/2041-8205/754/2/L29}, \href
  {https://ui.adsabs.harvard.edu/abs/2012ApJ...754L..29W} {754, L29}

\bibitem[\protect\citeauthoryear{{Whitaker} et~al.,}{{Whitaker}
  et~al.}{2014}]{whitaker14}
{Whitaker} K.~E.,  et~al., 2014, \mn@doi [\apj] {10.1088/0004-637X/795/2/104},
  \href {https://ui.adsabs.harvard.edu/abs/2014ApJ...795..104W} {795, 104}

\bibitem[\protect\citeauthoryear{{Wise}, {Turk}, {Norman}  \& {Abel}}{{Wise}
  et~al.}{2012}]{wise2012}
{Wise} J.~H.,  {Turk} M.~J.,  {Norman} M.~L.,   {Abel} T.,  2012, \mn@doi
  [\apj] {10.1088/0004-637X/745/1/50}, \href
  {https://ui.adsabs.harvard.edu/abs/2012ApJ...745...50W} {745, 50}

\bibitem[\protect\citeauthoryear{{Wisnioski} et~al.,}{{Wisnioski}
  et~al.}{2015}]{wisnioski15}
{Wisnioski} E.,  et~al., 2015, \mn@doi [\apj] {10.1088/0004-637X/799/2/209},
  \href {https://ui.adsabs.harvard.edu/abs/2015ApJ...799..209W} {799, 209}

\bibitem[\protect\citeauthoryear{{Wisnioski} et~al.,}{{Wisnioski}
  et~al.}{2019}]{wisnioski19}
{Wisnioski} E.,  et~al., 2019, \mn@doi [\apj] {10.3847/1538-4357/ab4db8}, \href
  {https://ui.adsabs.harvard.edu/abs/2019ApJ...886..124W} {886, 124}

\bibitem[\protect\citeauthoryear{{Wuyts} et~al.,}{{Wuyts}
  et~al.}{2011}]{wuyts11}
{Wuyts} S.,  et~al., 2011, \mn@doi [\apj] {10.1088/0004-637X/742/2/96}, \href
  {https://ui.adsabs.harvard.edu/abs/2011ApJ...742...96W} {742, 96}

\bibitem[\protect\citeauthoryear{{van Donkelaar}, {Agertz}  \& {Renaud}}{{van
  Donkelaar} et~al.}{2022}]{floor22}
{van Donkelaar} F.,  {Agertz} O.,   {Renaud} F.,  2022, \mn@doi [\mnras]
  {10.1093/mnras/stac692}, \href
  {https://ui.adsabs.harvard.edu/abs/2022MNRAS.512.3806V} {512, 3806}

\makeatother
\end{thebibliography}
